\documentclass[]{aa}
\usepackage{txfonts}
\usepackage{times}
\usepackage{natbib}
\usepackage{graphicx}
\usepackage{epsfig,longtable,lscape}
\usepackage{epsfig}
\usepackage{rotating}
\bibpunct[]{(}{)}{;}{a}{}{,}

\def\XMM{{\em XMM--Newton}}

\def\Chandra{{\em Chandra}}
\def\pn{{\em pn}}

\def\BDone{BD\,+37$^\circ$\,442}   
\def\BDtwo{BD\,+37$^\circ$\,1977}   
\def\BDthree{{BD+28$^\circ$ 4211}}

\def\HD{HD\,49798}
\def\approxgt{\mathrel{\hbox{\rlap{\lower.55ex \hbox {$\sim$}}
        \kern-.3em \raise.4ex \hbox{$>$}}}}
\def\approxlt{\mathrel{\hbox{\rlap{\lower.55ex \hbox {$\sim$}}
        \kern-.3em \raise.4ex \hbox{$<$}}}}

\def\mdot{\dot M}
\def\lx{$L_{\rm X}$}
\def\fx{$f_{\rm X}$}
\def\flux{\mbox{erg cm$^{-2}$ s$^{-1}$}}
\def\lum{\mbox{erg s$^{-1}$}}
\def\countsec{\hbox{counts s$^{-1}$}}

\def\nh{$N_{\rm H}$}
\def\ltsima{$\; \buildrel < \over \sim \;$}
\def\lsim{\lower.5ex\hbox{\ltsima}}
\def\gtsima{$\; \buildrel > \over \sim \;$}
\def\gsim{\lower.5ex\hbox{\gtsima}}
\def\msole{~M_{\odot}}

\def\chisqnu {$\chi^{2}_{\nu}$}

\begin{document}

\title{Follow-up observations of X-ray emitting hot subdwarf stars: the compact He-poor sdO star Feige 34\thanks{Based on observations obtained with \XMM, an ESA science mission with instruments and contributions directly funded by ESA Member States and NASA}}

\author{N. La Palombara\inst{1}, S. Mereghetti\inst{1}, P. Esposito\inst{1}, A. Tiengo\inst{1,2,3}}

\institute{INAF, Istituto di Astrofisica Spaziale e Fisica Cosmica - Milano, via A. Corti 12, I-20133, Milano, Italy
\and IUSS-Istituto Universitario di Studi Superiori, piazza della Vittoria 15, I-27100 Pavia, Italy
\and Istituto Nazionale di Fisica Nucleare, Sezione di Pavia, via A. Bassi 6, I-27100 Pavia, Italy}

\titlerunning{\XMM\ observation of Feige 34}

\authorrunning{La Palombara et al.}

\abstract{We report on results obtained with the \XMM\ observation of Feige 34 carried out in April 2018. This is the first spectroscopic X-ray observation of a compact and helium-poor hot subdwarf star. The source was detected at a flux level \fx\ = 3.4$\times10^{-14}$ erg cm$^{-2}$ s$^{-1}$ in the energy range 0.2--3 keV, which implies an X-ray-to-bolometric flux ratio $f_{\rm X}/f_{\rm bol} \simeq 10^{-6.5}$. The source spectrum can be described with the sum of two thermal-plasma components with subsolar abundances at temperatures of $\simeq$ 0.3 and 1.1 keV. These properties are similar to what is observed in early-type main-sequence stars, where the X-ray emission is attributed to turbulence and shocks in the stellar wind. Therefore, the same phenomenon could explain the X-ray properties of Feige 34. However, it is not possible to reproduce the observed spectrum with a thermal-plasma model if the elemental abundances are fixed at the values obtained from the optical and UV spectroscopy. Moreover, we show that the X-ray luminosity and spectrum are consistent with those expected from a young main-sequence star of late spectral type. Therefore, we discuss the possibility that the observed X-ray emission is due to the companion star of M0 spectral type, whose presence is suggested by the IR excess in the spectral energy distribution of Feige 34.
\keywords{stars: early-type --- stars: subdwarfs --- stars: individual: Feige 34 --- X-rays: stars}}

\maketitle

\section{Introduction}\label{introduction}

Hot subdwarf stars are subluminous blue stars that, in the Hertzprung-Russell (HR) diagram, lie between the main sequence and the white-dwarf (WD) sequence, at the blue end or beyond the horizontal branch (HB). They are the progeny of low-mass ($\sim 1 \msole$) main-sequence stars that have lost  most of their hydrogen envelopes during the red-giant phase, and are now burning their helium-rich core (see \citealt{Heber16} for a review). They are found in both the thin and the thick discs, and in the bulge and halo populations of the Galaxy \citep{Altmann+04,Busso+05,Geier+17}.

Based on their effective temperature, hot subdwarf stars are spectroscopically classified as either sdB, with T$_{\rm eff} \la$ 38 kK, or sdO, with T$_{\rm eff} \ga$ 38 kK \citep{Hirsch+08}. The class of the sdB stars is  homogeneous and most of them are helium poor, with  only weak helium lines or none at all. On the other hand, the sdO stars form a heterogeneus group; they display a wide range of effective temperatures (T$_{\rm eff}$ = 38--100 kK), surface gravities (log($g$)(cm s$^{-2}$) = 4--6.5), and helium abundances (--3.5 $\lsim$ log($N_{\rm He}/N_{\rm H}$) $\lsim$ 3) \citep{HeberJeffery92,Heber+06,Stroeer+07}. Therefore, depending on the atmospheric helium abundance, sdO stars are usually classified as either He poor or He rich. In addition, depending on the surface gravity, they can be classified as either luminous or compact \citep{Napiwotzki08}.

The variety among hot subdwarf stars is mostly due to different evolutionary histories. The sdB stars belong to the extreme horizontal branch (EHB) stars \citep{Heber86}. Since their hydrogen envelope is too thin to sustain hydrogen burning, after the exhaustion of the helium core they do not ascend the asymptotic giant branch (AGB), but evolve directly to the white-dwarf cooling sequence. The luminous sdO stars (both He poor and He rich) are post-AGB stars, while the compact He-poor sdO stars are post-EHB stars that very probably descend from the sdB stars. On the other hand, the origin of the compact He-rich sdO stars can be due to either the so-called late hot-flasher scenario \citep{Brown+01} or the merger of two WDs \citep{Iben90,SaioJeffery00,SaioJeffery02}.

Hot subdwarf stars are bright in optical and UV wavelength ranges, and are usually investigated in these particular spectral regions. In recent years, the high sensitivity of the instruments on board the \XMM\ and \Chandra\ space telescopes have allowed us to study the X-ray emission of this type of stars (see \citealt{Mereghetti&LaPalombara16} for a review).

In the past, our team used \XMM\ to perform deep observations of the three luminous and He-rich sdO stars \HD\ \citep{Mereghetti+13}, \BDtwo\ \citep{LaPalombara+15}, and \BDone\ \citep{LaPalombara+12,Mereghetti+17}. In all three cases the observed source spectrum can be described with the sum of multi-temperature thermal-plasma components (assuming the elemental abundances derived from the optical observations of these stars) with temperatures between $\simeq$ 0.1 and $\simeq$ 5 keV. The same type of spectrum has also been found  in a large sample of normal O-type stars observed with \XMM\ \citep{Naze09}. Moreover, for these three sdO stars the ratio of the X-ray to bolometric luminosities agrees with the `canonical' relation $L_{\rm X} \sim 10^{-7} \times L_{\rm bol}$, which has long been known for the main-sequence, giant, and supergiant O-type stars \citep{Pallavicini+81,Sciortino+90,GuedelNaze09}. The strong winds of these stars are characterized by turbulence phenomena and shock episodes that generate the observed X-ray emission \citep{Sundqvist+12a,Sundqvist+12b,Owocki+13,Cohen+14}. Compared to O-type stars, the bolometric luminosity of sdO stars detected in X-rays is significantly lower (log($L_{\rm bol}/L_{\odot}$) $\simeq$ 4 instead of 5--6). However, they have winds with mass-loss rates up to 10$^{-8}$ M$_{\odot}$ y$^{-1}$ \citep{Hamann10,JefferyHamann10} that can produce X-ray emitting shocks, as in more luminous O-type stars. This suggests that the X-ray emission of the sdO stars has the same origin as that observed in normal O-type stars.

The three sources discussed above are luminous He-rich sdO stars with a low surface gravity (log($g$) $\simeq$ 4), for which evidence of mass loss has been reported. However, our \Chandra\ programme of snapshot observations \citep{LaPalombara+14} allowed us to detect  Feige 34 and \BDthree\ as well; they  are compact He-poor subdwarfs with high surface gravity (log($g$) $>$ 6) and no sign of mass loss (see e.g. \citealt{Latour+13}). For these two stars $L_{\rm X} \sim 10^{-7} \times L_{\rm bol}$, as is true for  main-sequence stars, which suggests that the observed X-ray emission comes from the shock-heated gas in the stellar winds in their case as well.

In this paper we report on a follow-up observation of Feige 34, performed in April 2018 with \XMM, which allowed us to investigate in detail the spectral and timing properties of the X-ray emission discovered with \Chandra. Feige 34 is a bright ($V$ = 11.14) and well-known He-poor sdO star,  used as a standard star for flux calibration. Based on the results provided by the \textit{Gaia DR2}, it is at a distance $d = 226 \pm 5$ pc \citep{Bailer-Jones+18}. Very recently \citet{Latour+18} used high-quality optical and UV spectra (obtained with IUE\footnote{International Ultraviolet Explorer} and FUSE\footnote{Far Ultraviolet Spectroscopic Explorer}) to perform a comprehensive spectroscopic analysis of this star. They considered non-local thermodynamic equilibrium (NLTE) model atmospheres to estimate the fundamental atmospheric parameters and the elemental abundances. First, they  simultaneously fitted  the optical H and He lines to estimate the surface gravity, the effective temperature, and the He abundance, obtaining log($g$) $\simeq$ 6.0, T$_{\rm eff} \simeq$ 62 kK, and log($N_{\rm He}/N_{\rm H}$) $\simeq$ --1.8. The atmospheric parameters were then kept fixed, and the UV spectra were used to derive log($g$) = 5.99$\pm$0.03, T$_{\rm eff}$ = 62550$\pm$600 K, and the metal abundances that we report in Table~\ref{abundances}.

\begin{table}[!t]
\caption{Elemental abundances of Feige 34 relative to hydrogen (left column, \citealt{Latour+18}) and to solar abundances (right column).}\label{abundances}
\begin{center}
\begin{tabular}{ccc} \hline \hline
Element &       log[$N$(X)/$N$(H)]$_*$  &       [$N$(X)/$N$(H)]$_*$/[$N$(X)/$N$(H)]$_\odot$     \\ \hline
He              &       --1.79                                  &       0.166                                                                           \\
C               &       $\lsim$ --6.7                   &       $\lsim$ 8.3$\times 10^{-4}$                                        \\
N               &       --4.9                                   &       0.166                                                                           \\
O               &       --5.5                                   &       6.5$\times 10^{-3}$                                                        \\
Mg              &       $\lsim$ --5.0                   &       $\lsim$ 0.398                                                           \\
Si              &       --6.2                                   &       0.034                                                                           \\
P               &       --6.7                                   &       0.759                                                                           \\
S               &       --5.6                                   &       0.204                                                                           \\
Cr              &       $\lsim$ --5.3                   &       $\lsim$ 15.47                                                           \\
Mn              &       $\lsim$ --5.6                   &       $\lsim$ 11.47                                                           \\
Fe              &       --3.1                                   &       29.53                                                                           \\
Co              &       $\lsim$ --5.8                   &       $\lsim$ 19.05                                                           \\
Ni              &       --4.0                                   &       28.23                                                                           \\ \hline
\end{tabular}
\end{center}
\end{table}

The analysis of the photometric data of Feige 34, performed by \citet{Latour+18}, showed an IR excess, which implies the presence of a cool companion star. The estimated temperature of this star, obtained by fitting the spectral energy distribution (SED), is T$_{\rm eff}$ = 3848$^{+214}_{-309}$ K, which corresponds to a star of M0 spectral type. The surface ratio between the sdO and its cool companion derived from the SED fitting is consistent with both stars being at the same distance. However, the measured radial velocity of Feige 34, based on the results of the spectral analysis, is RV = 11.0$\pm$7.7 km s$^{-1}$, with no evidence of significant variations; this is in agreement with previous results \citep{Maxted+00,Han+11}. This finding can be explained if the sdO+M0 binary system has a long period and/or a low inclination.

\section{Observation and data analysis}\label{data}

\begin{table*}
\caption{Summary of the \XMM\ observation of Feige 34 (ID 0800100101).}\label{observation}
\begin{center}
\begin{tabular}{ccccccc} \\ \hline
EPIC    & Camera                & Camera        & Camera Time  & Net Exposure Time       & Extraction Radius     & Net Count Rate        \\
Camera  & Filter        & Mode          & Resolution   & (ks)                    & (arcsec)              & ($\times$ 10$^{-3}$\countsec)           \\ \hline
\pn\    & Medium        & Full Frame    & 73 ms        & 53.8                    & 15                    & 15.7$\pm$0.6         \\
MOS1    & Medium        & Full Frame    & 2.6 s        & 55.7                    & 15                    & 3.3$\pm$0.3         \\
MOS2    & Medium        & Full Frame    & 2.6 s        & 55.7                    & 15                    & 3.1$\pm$0.3         \\ \hline
\end{tabular}
\end{center}
\end{table*}

\begin{figure*}[]
\centering
\includegraphics[angle=0,width=18truecm]{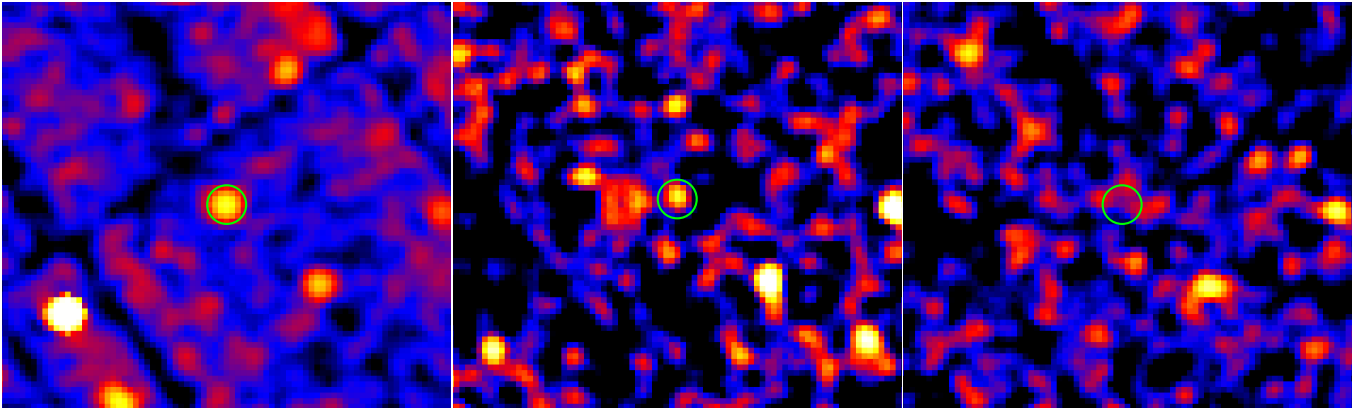}
\caption{Mosaic smoothed image of the three EPIC cameras of the sky region around Feiger 34 in the energy ranges 0.2-0.3 (\textit{left}), 3-4 (\textit{centre}), and 4-5 keV (\textit{right}). The source position is represented with a green circle of 15$''$ radius.}
\label{image}
\end{figure*}

Feige 34 was observed with \XMM\ on April 28, 2018 (MJD 58236), for a total exposure time of $\simeq$ 55 ks. The source flux is too low for a meaningful analysis of the data collected by the two Reflection Grating Spectrometers (RGS, \citealt{denHerder+01}). Therefore, we considered only the EPIC \pn\ \citep{Struder+01} and MOS \citep{Turner+01} focal-plane cameras; in Table~\ref{observation} we provide the set-up of these cameras. The event files of the three cameras were processed using version 16 of the \XMM\ \textit{Science Analysis System}\footnote{https://xmm-tools.cosmos.esa.int/external/xmm\_user\_support/documentation/sas\_usg/USG/} (\texttt{SAS}). We verified that the whole observation was characterized by a low instrumental background and that there was no contamination due to soft protons. This allowed us to consider the full set of EPIC data for our analysis. As shown in Fig.~\ref{image}, Feige 34 was detected down to 0.2 keV and up to 4 keV, while it remained undetected at higher energies.

For the \pn\ camera we selected events with pattern between 0 and 4 (corresponding to mono- and bi-pixel events), while for the two MOS cameras we considered events with patterns between 0 and 12 (corresponding to events involving between 1 and 4 pixels). We selected the same source and background extraction regions for the three cameras. The source events were extracted from a circular region centred at the source position and with a small radius of 15$''$ (to minimize the background contribution), while we accumulated the background events from a circular area free of sources and with a radius of 60$''$. In Table~\ref{observation} we list the corresponding source net count rates (CRs).

\begin{figure}
\begin{center}
\resizebox{\hsize}{!}{\includegraphics[angle=-90,clip=true]{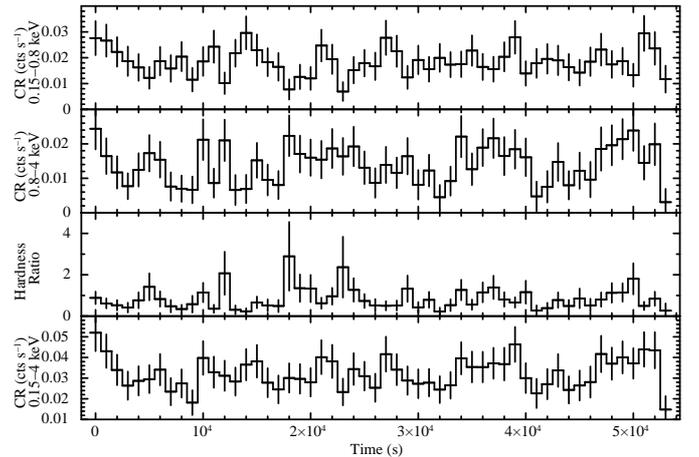}}
\caption{Background-subtracted light curves of Feige 34 in the energy ranges 0.15--0.8, 0.8--4, and 0.15--4 keV, with a time binning of 1000 s.}
\label{lc}
\end{center}
\vspace{-0.75 cm}
\end{figure}

For each camera we accumulated a light curve, with a time binning of 1000 s, in the three energy ranges 0.15-0.8 (soft), 0.8-4 (hard), and 0.15-4 keV (total). Then we used the \texttt{SAS} tool \textsc{epiclccorr} to correct each curve for both the background signal and the extraction region. In this way we obtained an  average CR in the total range for the \pn, MOS1, and MOS2 cameras of $\simeq$ 23.0$\times 10^{-3}$, 5.4$\times 10^{-3}$, and 4.0$\times 10^{-3}$ cts s$^{-1}$, respectively. Finally, for each of the three energy  ranges we summed the light curves of the individual cameras to obtain the cumulative EPIC light curve. The three curves are shown in Fig.~\ref{lc}, where we also show  the hardness ratio of the hard (H) to the soft (S) light curves (HR = H/S). Each curve shows no significant flux increase or decrease along the whole observation: we found no evidence of flux or spectral variability since a fit with a constant is fully acceptable for both the CR and the HR. The average CR in the soft, hard, and total range is $\simeq$ 1.85$\times 10^{-2}$, 1.39$\times 10^{-2}$, and 3.24$\times 10^{-2}$ \countsec, respectively.

%%%%%%%%%%%%%%%%%%%%%%%%%%%%%%%%%%%%%%%%%%%%%%%%%%%%%%%%%%%%%%%%%%%%%%%% 
%%%%%%%%%%%%%%%%%%%%%%%%%%%%%%%%%%%%%%%%%%%%%%%%%%%%%%%%%%%%%%%%%%%%%%%%

Since the light curve reported in Fig.~\ref{lc} shows no signs of significant flux or spectral variability along the observation, we considered the whole exposure for the source spectral analysis. We accumulated the source spectrum for each of the three EPIC cameras, using the same extraction regions considered for the light curves. However, due to the low source flux, the {signal-to-noise} ratio of these spectra was very low. Therefore, we used the \texttt{SAS} task \textsc{epicspeccombine} to combine them in a single spectrum and to calculate the applicable response matrix and ancillary file. We rebinned the spectrum with a minimum of 30 net counts per bin and performed the spectral analysis in the energy range between 0.2 and 3 keV (since there was no significant bin at higher energies). The rebinned spectrum was fitted using version 12.9.1 of \texttt{XSPEC} and the spectral uncertainties were calculated at the 90 \% confidence level for one interesting parameter. For the spectral fitting we used the absorption model \textsc{tbabs} in
\texttt{xspec}. We considered the photoelectric absorption cross sections of \citet{Verner+96}, and we adopted the results of \citet{WilmsAllenMcCray00} for the elemental abundances.

\begin{figure}[h]
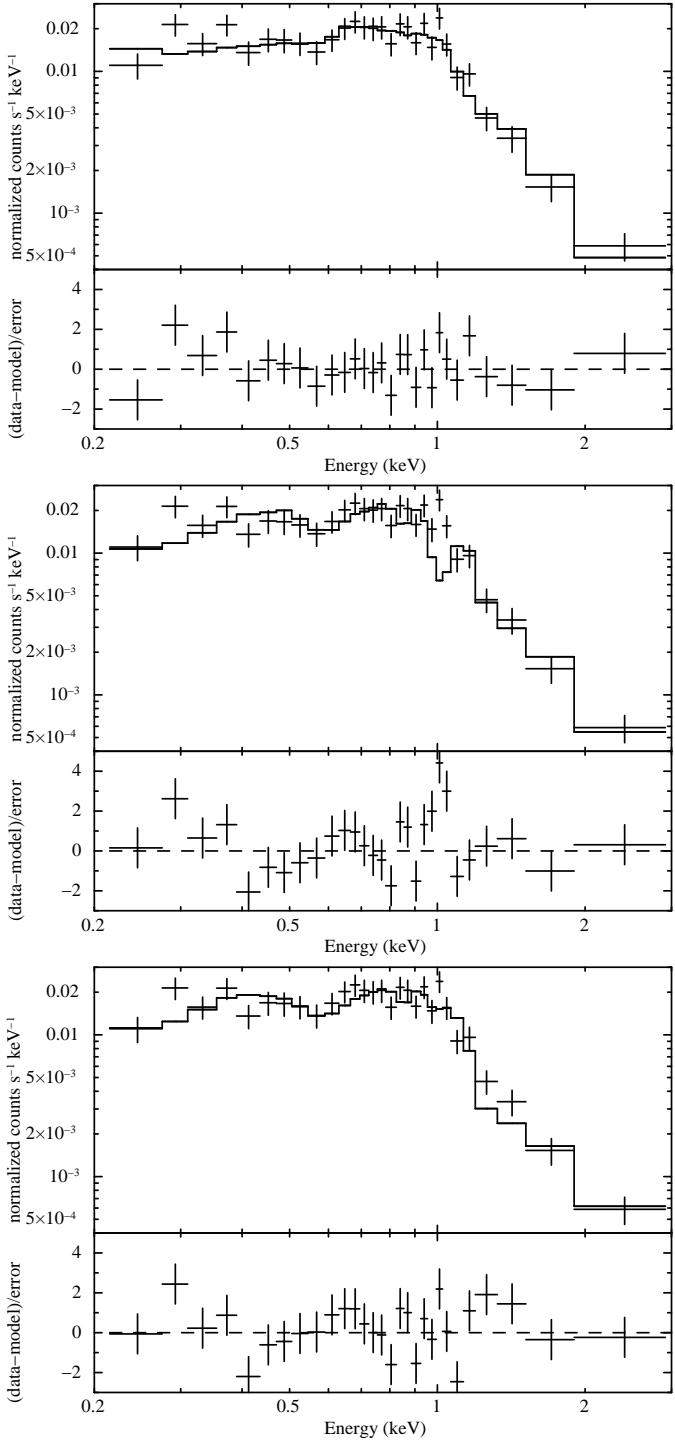

\centering
\resizebox{\hsize}{!}{\includegraphics[angle=-90,clip=true]{tbabs_2apec_afree_new.ps}} 

\resizebox{\hsize}{!}{\includegraphics[angle=-90,clip=true]{tbabs_2vvapec-Latour18-corretta_new.ps}} 

\resizebox{\hsize}{!}{\includegraphics[angle=-90,clip=true]{tbabs_2vvapec-Latour18-corretta_Si-free.ps}} 
\caption{EPIC spectrum of Feige 34 with the best-fit model composed of the sum of (1) two \textsc{apec} components with free elemental abundances (\textit{top panel}), (2) two \textsc{vvapec} components with elemental abundances fixed at the values estimated by \citet{Latour+18} (\textit{middle panel}), and (3) the same as in (2) but with free Si abundance (\textit{bottom panel}).}
\label{spectrum}
\end{figure}

The EPIC spectrum of Feige 34 is very soft (see Fig.~\ref{spectrum}), and using single-component models the best fit was obtained with an \textsc{apec} model (which represents the spectral emission due to a collisionally ionized gas) with abundance $\sim$ 0.1 solar (\chisqnu\ = 1.76). We tried to describe the spectrum with the sum of a power-law and a thermal component (either a black body or a bremsstrahlung) or to improve the fit with the \textsc{apec} model by considering additional components, but in all cases the best-fit parameters were unconstrained or had unrealistic values. Therefore, as in the case of the other sdO stars we investigated with \XMM, we tried to describe the spectrum of Feige 34 with the sum of two \textsc{apec} components at different temperatures. We obtained an acceptable fit (\chisqnu\ = 1.20) by leaving the abundance free to vary (Fig.~\ref{spectrum}, top panel). In Table~\ref{spectral_fit} we list the best-fit parameters obtained for this spectral model. We tried to improve the spectral fit with the addition of a third \textsc{apec} component, but this attempt was unsuccessful: the \chisqnu\ did not reduce and, moreover, the additional component was unconstrained.

\begin{table}[!t]
\caption{Best-fit parameters of Feige 34 obtained with the sum of two thermal plasma emission models (\textsc{apec}) and free element abundance.}\label{spectral_fit}
\begin{center}
\begin{tabular}{ccc} \hline \hline
Parameter                                                       & Unit                           & Value                 \\ \hline
\nh                                                                     & cm$^{-2}$                       & (2.2$^{+2.0}_{-1.6}$)$\times10^{20}$  \\
Abundance                                                               & -                                                       & 0.21$^{+0.24}_{-0.09}$                                \\
$kT_{APEC1}$                                                    & keV                 & 0.30$^{+0.07}_{-0.05}$                \\
Flux$_{\rm APEC1}$(0.2-3 keV)$^{(a)}$   & $\times 10^{-14}$ \flux       & 1.6$^{+0.5}_{-0.4}$                                     \\
$kT_{APEC2}$                                                    & keV                                           & 1.1$^{+0.2}_{-0.1}$                                     \\
Flux$_{\rm APEC2}$(0.2-3 keV)$^{(a)}$   & $\times 10^{-14}$ \flux       & 1.8$^{+0.4}_{-0.3}$                                     \\
Flux$_{\rm TOT}$(0.2-3 keV)$^{(a)}$             & $\times 10^{-14}$ \flux       & 3.4$^{+0.5}_{-0.4}$                                     \\
Luminosity (0.2-3 keV)$^{(b)}$                  & $\times 10^{29}$ \lum         & 2.0$^{+0.2}_{-0.3}$                                     \\
\chisqnu/d.o.f.                                         & -                                             & 1.20/23                                                                 \\ \hline
\end{tabular}
\end{center}
$^{(a)}$ Corrected for absorption
\\
$^{(b)}$ Assuming a source distance $d$ = 226 pc
\end{table}

In our spectral analysis we also took into account  the results obtained by \citet{Latour+18}, who estimated the abundance of several elements of Feige 34 through the analysis of high-quality optical and UV spectra. To this end, we tried to describe the source spectrum with a multi-temperature thermal plasma model characterized by elemental abundances equal to those found by \citet{Latour+18}. Therefore, we considered a model composed of the sum of two \textsc{vvapec} components at different temperatures, where the hydrogen abundance was fixed to 1 and the abundances of the heavier elements were fixed to the values (relative to solar) given in Table~\ref{abundances}. We found that it is not possible to describe the source spectrum with this type of model since it is fully rejected by our data. As shown in Fig.~\ref{spectrum} (middle panel), the best-fit model obtained in this way fails to reproduce the observed spectrum since it leaves large residuals over a wide energy range (especially at E $\sim$ 1 keV). This is reflected in the high value of the reduced chi-squared (\chisqnu\ = 2.7). We tried to improve the spectral fit with this type of model by leaving  the abundance of single elements free to vary. We first considered the Ne abundance in order to reproduce the feature at $\sim$ 1 keV. We found that the Neon abundance is well constrained (1.5$^{+0.8}_{-0.6}$ compared to the solar value), but the fit improvement was very limited and, moreover, large residuals around 1 keV were still present. Instead, we obtained a considerably better result by leaving  the abundance of either Si or S free to vary. In these cases the value of \chisqnu\ reduced to 1.83 and 2.00, respectively, and the spectral feature at $\simeq$ 1 keV disappeared. However, the best-fit value of the element abundance is very high (120$^{+70}_{-50}$ and 115$^{+100}_{-50}$ for Si and S, respectively), and is thus  unreasonable. Moreover, in both cases the fit still leaves large residuals at various energies (see Fig.~\ref{spectrum}, bottom panel, for the Si case). When left free to vary, it was also possible to constrain the abundance value  for O, Ar, Cr, Mn, and Fe. In none of these cases, however, did the fit improve  significantly compared to the model with the abundances fixed at the values estimated by \citet{Latour+18}. This means that, with this type of model, it was never possible to obtain a  goodness of fit comparable to that provided by the model given in Table~\ref{spectral_fit}.

\section{Discussion}

The timing analysis of the \XMM\ data shows no evidence of flux or spectral variability along the observation. The 0.2-3 keV source flux is $f_{\rm X} = 3.4 \times 10^{-14}$ erg cm$^{-2}$ s$^{-1}$, which agrees with the result obtained with \Chandra. For the source distance $d$ = 226 pc, this flux implies an X-ray luminosity $L_{\rm X} \simeq 2 \times 10^{29}$ erg s$^{-1}$. The source spectrum can be described with a combination of two thermal-plasma emission components at different temperatures (a few MK), provided that a subsolar elemental abundance is used. This type of emission model is commonly used to describe the X-ray spectrum of normal O-type stars, which are a well-known class of X-ray sources with X-ray luminosities up to \lx\ $\sim 10^{33}$ erg s$^{-1}$. They are characterized by strong radiatively driven stellar winds, with a clumped structure and mass-loss rates up to $\mdot_{\rm W} = 10^{-5}$ M$_{\odot}$ y$^{-1}$. The clump--clump collisions heat up the cool material, thus creating a hot plasma at $T \sim$ 1-10 MK which generates the observed X-ray emission. In the case of Feige 34 the high surface temperature favours the presence of a radiatively driven wind, although the predicted mass-loss rate is much lower than in normal O-type stars ($\mdot_{\rm W} \simeq 10^{-7.5}$ M$_{\odot}$ y$^{-1}$ according to \citealt{Thejll+95}, and even down to $\mdot_{\rm W} \simeq 10^{-10}$ M$_{\odot}$ y$^{-1}$ according to \citealt{krticka+16}). Therefore, it is possible that  for this sdO star the observed X-ray emission has the same origin.

Since for normal O-type stars $f_{\rm X}/f_{\rm bol} = 10^{-6.7}$ \citep{Naze09}, it is interesting to compare the X-ray and the bolometric fluxes of Feige 34. To this end, we estimated $f_{\rm bol}$ based on the $V$ magnitude, the interstellar reddening (A$_V$), and the bolometric correction (BC) of the star. Since E($B-V$) = 0.018 \citep{Latour+18}, the relation A$_V$ = 3.2$\times$E($B-V$) \citep{Zombeck} implies A$_V$ = 0.0576. Then, assuming T$_{\rm eff}$ = 62,550 K \citep{Latour+18}, from the relation BC = 27.66 - 6.84$\times$log(T$_{\rm eff}$) \citep{Vacca+96} we obtained BC = -5.15. Since $V$ = 11.14 \citep{Hog+00} and m$_{bol}$ = $V$ - A$_V$ + BC \citep{Zombeck}, we obtained m$_{bol}$ = 5.93. Finally, from the relation $f_{bol}$ = 2.48$\times10^{-5}\times10^{-0.4{\rm m}_{bol}}$ \citep{Zombeck}, we obtained $f_{bol}$ = 1.05$\times10^{-7}$ \flux. This implies log($f_{\rm X}$/$f_{\rm bol}$) = -6.48, a result near the average value obtained for the normal O-type stars. Therefore,  the X-ray-to-bolometric flux ratio also supports the hypothesis that the X-ray emission observed in Feige 34 originates from the hot plasma in the stellar wind.

In Table~\ref{parameters} we compare the main parameters of Feige 34 with those of the three sdO stars already observed with \XMM. Contrary to Feige 34, these stars are luminous and He rich and have a very different chemical composition. Moreover, they are characterized by much lower temperatures and surface gravities and higher mass-loss rates. Compared to Feige 34, they have much higher X-ray and bolometric luminosities\footnote{The values of the bolometric luminosity listed in Table~\ref{parameters} are taken from the values published in the literature, rescaled for the distances based on Gaia results}. However, since  their distances (estimated on the basis of the \textit{Gaia DR2}) are also significantly  larger, their X-ray fluxes are very similar to that observed for Feige 34. The X-ray spectra of these stars can be described with the sum of different thermal-plasma components at various temperatures, if the specific elemental abundances are properly taken into account. The temperatures of these components, however, are lower than those of Feige 34, which implies that the X-ray spectra are softer.

\begin{table*}[t]
\caption{Main parameters of the sdO stars Feige 34, \BDtwo, \BDone, and \HD.}\label{parameters}
\begin{center}
\begin{tabular}{l|cc|cc|cc|cc} \hline \hline
Parameter                       & \multicolumn{2}{c|}{Feige 34} & \multicolumn{2}{c|}{\BDtwo}     & \multicolumn{2}{c|}{\BDone}           & \multicolumn{2}{c}{\HD}                    \\
                                & Value         & Reference             & Value         & Reference       & Value                 & Reference     & Value                 & Reference       \\ \hline
log $g$ (cm s$^{-2}$)                   & 5.99                  & 1                              & $\simeq$ 4.0  & 6               & 4.00 $\pm$ 0.25       & 11            & 4.35                  & 14               \\
$T_{\rm eff}$ (K)                               & 62,550                & 1                               & 48,000        & 7               & 48,000                & 7             & 46,500                & 14               \\
$U$                                             & 9.61                  & 2                               & 8.67          & 8               & 8.57                  & 12            & 6.76                  & 15               \\
$B$                                             & 10.91                 & 2                               & 9.93          & 8               & 9.73                  & 12            & 8.02                  & 15               \\
$V$                                             & 11.14                 & 2                               & 10.17         & 8               & 10.01                 & 12            & 8.29                  & 15               \\
$d_{\rm Gaia}$ (pc)                     & 226$\pm$5             & 3                             & 1,200$^{+180}_{-140}$ & 3       & 1,230$^{+320}_{-220}$ & 3             & 501$^{+17}_{-16}$     & 3               \\
$L_{\rm Gaia}^{(a)}$ ($L_{\odot}$) & 158        & 4                             & 4,900 & 6   & 9,500 & 6         & 8,300 & 14               \\
v$_{\rm W}$ (km s$^{-1}$)               & -                             & -                               & 2,000         & 7               & 2,000                 & 7             & 1,200                 & 9               \\
$\dot M_{\rm W}$ ($M_{\odot}$ yr$^{-1}$)& 10$^{-10}$    & 5             & 10$^{-8.2}$   & 7               & 10$^{-8.5}$           & 7             & 10$^{-9.2}$           & 9                       \\
\nh\ ($\times10^{20}$ cm$^{-2}$)   &  2.2       & 4     & 1     & 9     & 5       & 9     & 5     & 9     \\
kT$_1$  (keV)  & 0.30   & 4     & 0.13  & 9     & 0.11  & 9             & 0.11    & 9     \\
kT$_2$  (keV)  & 1.10   & 4     & 0.79  & 9     & 0.65  & 9             & 0.57    & 9     \\
kT$_3$  (keV)  & -              & 4     & -             &       & -             &                 & 4             & 9     \\
\fx$^{(b)}$ ($\times 10^{-14}$ \flux)   & 3.4$^{+0.5}_{-0.4}$   & 4     & 4.0$^{+0.2}_{-0.3}$     & 10    & 3.4$^{+0.3}_{-0.1}$   & 13            & 9.2$\pm$0.7                                     & 16                                    \\
\lx$^{(c)}$ ($\times 10^{30}$ \lum)     & 0.20$^{+0.04}_{-0.03}$        & 4                               & 6.5$^{+2.5}_{-1.8}$                   & 10                              & 5.8$^{+4.2}_{-2.0}$                                   & 13                              & 2.6$\pm$0.2                                   & 16                                      \\
log(\lx/$L_{\rm bol}$)$^{(d)}$                  & --6.48$\pm$0.06                       & 4                               & --6.46$^{+0.02}_{-0.03}$                      & 10                              & --6.80$^{+0.04}_{-0.01}$                                      & 13                              & --7.09$\pm$0.03                                       & 16                                      \\ \hline
Element & \multicolumn{8}{c}{Abundances$^{(e)}$} \\ \hline
$X_{\rm H}$             & 0.89663       & 1            &     0.0013  & 11      & 0.0013  & 11  &  0.19       & 14      \\
$X_{\rm He}$    & 0.05774       & 1      &     0.9639   & 11     &     0.9639  & 11     &  0.78  & 14       \\
$X_{\rm  C}$    & $\lsim$ 2.1$\times 10^{-6}$ & 1        &     0.0250  & 11    &     0.0250  & 11    &  0.0001     & 14       \\
$X_{\rm  N}$    & 0.00016       & 1        &      0.0031  & 11  &      0.0031  & 11  &  0.025  & 14    \\
$X_{\rm  O}$    & 4.5$\times 10^{-5}$   & 1         &      0.0053  & 11   &      0.0053  & 11   &  0.0028  & 14 \\
$X_{\rm  Mg}$    & $\lsim$ 0.00022      & 1         & - &            -  &       -       &       -       &       -       &       -       \\
$X_{\rm  Si}$    & 1.6$\times 10^{-5}$  & 1                 &       0.0008  & 11 &       0.0008  & 11 &  0.001  & 11   \\
$X_{\rm  P}$    & 5.5$\times 10^{-6}$   & 1         &        -  &       -       &       -       &       -       &       -       &       -       \\
$X_{\rm  S}$    & 7.2$\times 10^{-5}$   & 1         &        -  &       -       &       -       &       -       &       -       &       -       \\
$X_{\rm  Cr}$    & $\lsim$ 0.00023      & 1         &        -  &       -       &       -       &       -       &       -       &       -       \\
$X_{\rm  Mn}$    & $\lsim$ 0.00012      & 1         &        -  &       -       &       -       &       -       &       -       &       -       \\
$X_{\rm  Fe}$    & 0.03946                      & 1       &      0.0006  & 7 &      0.0006  & 7 &  0.0011  & 14  \\
$X_{\rm  Co}$    & $\lsim$ 8.3$\times 10^{-5}$  & 1         &   -       &       -       &       -       &       -       &       -       &       -       \\
$X_{\rm  Ni}$    & 0.00522      & 1         &        -  &       -       &       -       &       -       &       -       &       -       \\ \hline
    \hline
\end{tabular}
\end{center}
\begin{small}
Notes: $^{(a)}$ Based on the reported reference, corrected for the Gaia-estimated distance. $^{(b)}$ Unabsorbed flux in the energy range 0.2-10 keV. $^{(c)}$ The errors on the source X-ray luminosity take into account the errors on both the source flux and distance. $^{(d)}$ The errors on \lx/$L_{\rm bol}$ take into account only the errors on the source X-ray flux. $^{(e)}$ Mass fraction.
\\
References: 1 - \citet{Latour+18}; 2 - \citet{Hog+00}; 3 - \citet{Bailer-Jones+18}; 4 - This work; 5 - \citet{krticka+16}; 6 - \citet{Darius+79}; 7 - \citet{JefferyHamann10}; 8 - \citet{Jordi+91}; 9 - \citet{Mereghetti&LaPalombara16}; 10 - \citet{LaPalombara+15}; 11 - \citet{BauerHusfeld95}; 12 - \citet{Landolt73}; 13 - \citet{Mereghetti+17}; 14 - \citet{Hamann10}; 15 - \citet{LandoltUomoto07};  16 - \citet{Mereghetti+16}
\end{small}
\end{table*}

\begin{figure}[h]
\centering
\resizebox{\hsize}{!}{\includegraphics[angle=-90,clip=true]{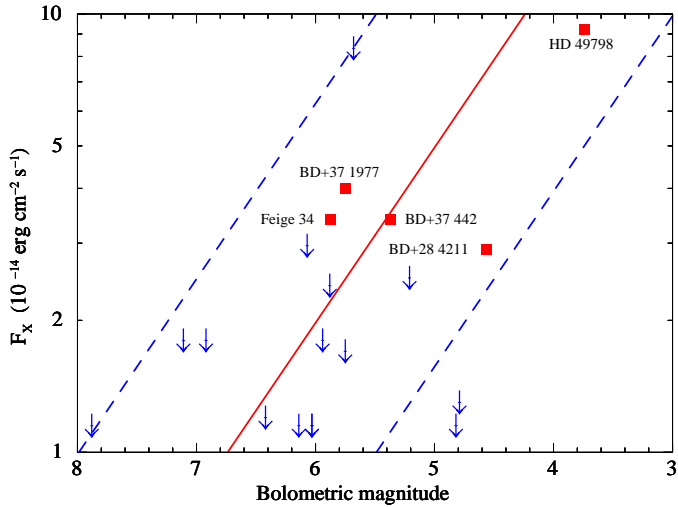}}
\caption{Relation between the X-ray flux (or its upper limit for the undetected sources) and the bolometric magnitude of the sdO stars observed at X-rays. The continuous red line represents the best-fit relation ($f_{\rm X}/f_{\rm bol} = 10^{-6.7}$) for the main-sequence early-type stars \citep{Naze09}, while the area between the two blue lines ($f_{\rm X}/f_{\rm bol} = 10^{-6.2}$ and $f_{\rm X}/f_{\rm bol} = 10^{-7.2}$, respectively) corresponds to the dispersion of this relation.}
\label{X-bol}
\end{figure}

In Fig.~\ref{X-bol} we put the four sdO stars investigated with \XMM\ in the context of the X-ray-observed sdO stars, thus also including  those studied with \Chandra. Both the X-ray flux of the detected stars and the upper limit of the undetected ones are reported as a function of their bolometric magnitude. The flux value of the four stars listed in Table~\ref{parameters} is obtained through the spectral fit of the \XMM\ data, while for the other sources we referred to the count rate value (or its upper limit) provided by \Chandra\ observations. For comparison, in the plot we also report  the average relation between X-ray and bolometric flux ($f_{\rm X}/f_{\rm bol} = 10^{-6.7 \pm 0.5}$) of the normal O-type stars \citep{Naze09}. The plot shows that the region delimited by this relation includes almost all the stars.

The results we obtained for the three sdO stars already observed with \XMM\ favoured the hypothesis that their X-ray emission is generated by shocks and turbulence in their winds \citep{LucyWhite80,Owocki+88}, as in the case of the normal O-type stars. Therefore, although Feige 34 differs from the other sdO stars in several ways, it is possible that the same emission scenario is also applicable  to this star; however, we note  an important difference. In the case of the other three sdO stars, the spectral fit with a multi-temperature thermal plasma model was obtained considering the specific abundance of each element obtained from the spectroscopic analysis in the optical/UV domain. The same approach was unsuccessful in the case of Feige 34; as shown in the previous section, it was not possible to obtain an acceptable spectral fit when we fixed the elemental abundances at the values obtained by \citet{Latour+18} from the optical data. On the other hand, a good-quality fit was only possible for  subsolar metallicity, with the relative elemental abundances kept solar.

It is also possible that  the X-ray emission detected with \XMM\ (or at least part of it)\  is due to the late-type companion (of M0 stellar type) of Feige 34, whose presence is inferred from the IR excess in the SED \citep{Thejll+95,Latour+18}. Main-sequence stars of late spectral types are established X-ray sources since the epoch of the \textit{Einstein} satellite \citep{Pallavicini+81,Vaiana+81}, and their X-ray emission is generally attributed to the effect of magnetic heating of the coronal plasma (at temperatures $ T >$ 1 MK). These stars have deep convective envelopes that  combine with the differential rotation to produce strong magnetic dynamos at the base of the convection zone. In turn, these dynamos produce high levels of magnetic activity above the stellar photosphere, which is traced by the coronal X-ray emission \citep{Guedel04}. As a result, their coronae are characterized by a mixture of cool ($T$ = 1.5--5 MK) and hot ($T$ = 10--30 MK) magnetic loops \citep{GuedelNaze09}. The observations of late-type stars performed with \textit{Einstein} showed that their spectra are due to the sum of two thermal components \citep{Vaiana83,Schmitt85,Majer+86}. In particular, \citet{Schmitt+90} found that for most stars a two-temperature (2T) model with $kT_1 \simeq$ 0.22 keV and $kT_2 \simeq$ 1.37 keV provides an adequate spectral fit. This finding was confirmed by the results of the \XMM\ Bright Serendipitous Survey (XBSS), which demonstrated that the spectrum of moderately active K and M stars can be fit using a 2T model with $kT_1$ = 0.32 keV and $kT_2$ = 0.98 keV \citep{LopezSantiago+07}.

The spectral analysis of the EPIC data of Feige 34 provided results that are consistent with the previous scenario. For this sdO star we need two thermal plasma components to obtain an adequate fit, and their temperatures (0.3 and 1.1 keV) are very similar to those of the M-type stars of the XBSS. Moreover, the estimated X-ray luminosity (\lx\ = 2$\times10^{29}$ \lum) is consistent with that typically observed in young M0 stars \citep{Pizzolato+03,Garces+11,Stelzer+13}. Regarding the X-ray-to-bolometric flux ratio, we estimated $L_{bol}$ assuming, for a main-sequence star of M0 type, that log($L_{bol}$/$L_{\odot}$) = --1.2 \citep{Allen73}. This implies $L_{bol} = 0.063 \times L_{\odot} = 2.4 \times 10^{32}$ \lum\ and, then log($L_{\rm X}/L_{\rm bol}$) = --3.1. This value agrees with the luminosity ratio observed in the most active stars, which saturates at a level log($L_{\rm X}/L_{\rm bol}$) $\simeq$ --3 \citep{Zickgraf+05,Stelzer+16,Kastner+16}.

\section{Conclusion}

The \XMM\ follow-up observation of Feige 34 performed in April 2018 allowed us to investigate with better data the properties of the X-ray emission detected with \Chandra\ in 2013 \citep{LaPalombara+14}. We showed that the source spectrum can be described with a combination of two thermal-plasma emission components at different temperatures, provided that a subsolar elemental abundance is used. The same type of model is used to describe the X-ray spectrum of normal O-type stars, where the X-ray emission is due to shocks and turbulence in the radiatively driven stellar winds. We already suggested this scenario for the three sdO stars previously observed with \XMM. Therefore, it is possible that it also applies  for Feige 34. This is also supported by the X-ray-to-bolometric flux ratio of Feige 34, which is fully consistent with that observed in early-type main-sequence stars. However, contrary to the other sdO stars observed with \XMM, we verified that it is not possible to obtain an acceptable spectral fit when the elemental abundances are fixed at the values obtained from the spectroscopic analysis of the optical and UV data.

The IR excess observed in the SED of Feige 34 suggests the presence of a late-type companion star of M0 spectral type. We show that the properties of the observed X-ray emission are consistent with those typical of young M-type stars. Therefore, although we cannot exclude that the observed X-ray emission originates in the sdO star itself, our results favour the possibility that its main source is the companion star. In this framework, we also note that the sdO star Feige 67, which is very similar to Feige 34 in several ways, remained undetected in our programme of snapshot observations of sdO stars performed with \Chandra\ \citep{LaPalombara+14}\footnote{where Feige 67 was identified as BD+18$^\circ$ 2647}, which had a sensitivity of $\simeq 10^{-14}$ \flux.

\begin{acknowledgements}
We acknowledge the financial contribution from the Italian Space Agency from ASI--INAF agreement n.\,2017-14-H.0.
PE acknowledges funding in the framework of the project ULTraS (ASI--INAF contract n.\,2017-14-H.0).
\end{acknowledgements}

\bibliographystyle{aa}
\bibliography{biblio}

\end{document}